\documentclass[aps,pre,reprint,groupedaddress,amsmath,amssymb,amsfonts,floatfix]{revtex4-2}

\usepackage{float}
\usepackage{graphicx}
\usepackage[caption=false]{subfig}

\usepackage[colorlinks,linkcolor=blue,anchorcolor=blue,citecolor=blue,urlcolor=blue]{hyperref}

\begin{document}


\title{Fluctuating resource availability stabilizes coexistence beyond the oscillatory tragedy of the commons}


\author{Yi-Duo Chen}

\author{Jian-Yue Guan}
\email[Corresponding author: ]{guanjy@lzu.edu.cn}

\author{Zhi-Xi Wu}
\email[Corresponding author: ]{wuzhx@lzu.edu.cn}

\affiliation{Lanzhou Center for Theoretical Physics, Key Laboratory of Theoretical Physics of Gansu Province, Key Laboratory of Quantum Theory and Applications of MoE, Gansu Provincial Research Center for Basic Disciplines of Quantum Physics, Lanzhou University, Lanzhou 730000, China}%



\begin{abstract}
Fluctuations in ambient conditions such as temperature and humidity can alter the effective availability of a shared resource at a given abundance. Standard evolutionary games with environmental feedback treat the payoff matrix as a deterministic function of resource abundance, leaving open how fluctuations in availability affect the oscillatory tragedy of the commons. Using noise-corrected replicator dynamics and individual-based birth-death simulations, we show that zero-mean fluctuations in resource availability can stabilize an interior equilibrium and suppress coupled strategy-resource oscillations. We derive an analytical criterion, in terms of the payoff structure and feedback dynamics, for when increasing noise favors local stability. These findings identify environmental variability as a mechanism that can promote strategy coexistence and help maintain shared resources in systems prone to recurrent depletion.
\end{abstract}


\maketitle

\section{Introduction}

The tragedy of the commons (TOC) is a central problem in coevolutionary game dynamics with environmental feedback \cite{hardin1968Tragedy,weitz2016oscillating,lin2019Spatial,tilman2020Evolutionary,chen2026reciprocityinduced,zhong2026spatial,chen2025coevolutionary,itoCompleteClassificationEvolutionary2024}. Unrestrained exploitation of a common resource by self-interested individuals inevitably depletes the resource and drives the entire system into a barren environmental state, a dilemma that, as Hardin argued, can be resolved only through moral restraints \cite{hardin1968Tragedy}. From the perspective of evolutionary game theory \cite{axelrod1981evolution,szabo2007evolutionary,nowak2006evolutionary,nowak2006five,nowak1992evolutionary}, however, mechanisms operating at multiple levels can support the evolution of cooperation rather than mutual defection \cite{weitz2016oscillating,lin2019Spatial,tilman2020Evolutionary,chen2026reciprocityinduced,zhong2026spatial,chen2025coevolutionary,itoCompleteClassificationEvolutionary2024}. Notably, the coupling of two payoff matrices with distinct feedback structures gives rise to oscillatory tragedy-of-the-commons (o-TOC) dynamics, characterized by alternating dominance of cooperators and defectors, together with environmental oscillations between abundance and scarcity \cite{weitz2016oscillating}.

In real-world complex systems, the effects of environmental conditions on individual evolution (i.e., payoff structure and transition probability) are not always as ideally deterministic as assumed in theoretical models. Fluctuations in ambient conditions can alter the rate or cost of acquiring shared resources \cite{dahake2022Signallike,demanincor2026Climate,domeignoz-horta2020Microbial,marshall2021Wholeorganism,smith2021Systematic,zheng2019Growth,bacheler2020catchability}. For example, mussels acclimated to fluctuating temperatures fed on algae more slowly than those held at a constant temperature \cite{marshall2021Wholeorganism}. In hawkmoth foraging experiments, floral humidity reduced the effort required to probe for nectar \cite{dahake2022Signallike}. In paired trap and video surveys of reef fish, trap catchability was associated with water temperature after accounting for local fish abundance \cite{bacheler2020catchability}. Such variation in resource access can translate into fluctuations in the payoffs of resource use.
Fluctuations in payoff structures can render evolutionary dynamics non-ergodic and alter the effective game type \cite{stollmeier2018Unfair}. Noise-induced replicator dynamics can systematically alter the stability structure of games and give rise to multistability \cite{wang2026Noiseinduced}. More broadly, parameter disorder can enhance dynamical stability in physical and engineered systems \cite{dsouza2026Noise,montanari2026Disorderpromoted}. 

Motivated by these studies, we ask how fluctuations in resource availability alter o-TOC dynamics and under what conditions they can stabilize coexistence. To address this question, we apply noise-corrected replicator dynamics \cite{wang2026Noiseinduced,taylor1978evolutionarya} to an evolutionary game with environmental feedback \cite{weitz2016oscillating,lin2019Spatial,tilman2020Evolutionary}. In our model, zero-mean fluctuations in resource availability enter through the payoff matrix. We derive an analytical criterion for when increasing noise favors local stability of the interior equilibrium, expressed in terms of the payoff structure and feedback dynamics. Individual-based birth-death simulations with Gaussian white noise in resource availability corroborate the predicted change in local stability and show that fluctuations can reduce the amplitude of coupled strategy-resource oscillations.


\section{Methods}

\subsection{Replicator dynamics}

Following the study in Ref.~\cite{weitz2016oscillating}, which identified the o-TOC dynamics, the temporal evolution of the environmental state $n \in [0, 1]$ is defined as 
\begin{equation}\label{eq:te_env}
    \dot{n} = \varepsilon n (1-n) [\theta x - (1-x)],
\end{equation}
where $x \in [0, 1]$ is the cooperator fraction, cooperators restore environmental resources at a rate of $\theta > 0$, whereas defectors consume environmental resources at a constant rate of $1$. Here, $\varepsilon > 0$ denotes the rate of environmental updating relative to strategy updating.

We consider fluctuations in resource availability and define the resource available to individuals as $\tilde{n} = n + \sigma \xi$, where $\xi$ denotes Gaussian white noise and $\sigma^2$ characterizes the variance of $\tilde{n}$. We allow $\tilde{n}$ to take arbitrary values to account for extreme fluctuations. Accordingly, the payoff matrix incorporating environmental feedback, as in Ref.~\cite{weitz2016oscillating}, is given by
\begin{equation}\label{eq:payoff_matrix_main}
    \begin{aligned}
        \Pi(\tilde{n}) &= \tilde{n} \Pi_1 + (1-\tilde{n}) \Pi_0 \\ 
    &=\tilde{n} 
    \begin{bmatrix}
        R_1 & S_1 \\
        T_1 & P_1 \\    
    \end{bmatrix} 
    + (1-\tilde{n}) 
    \begin{bmatrix}
        R_0 & S_0 \\
        T_0 & P_0 \\  
    \end{bmatrix}\\
    &=n\Pi_1 + (1-n)\Pi_0 + \sigma \xi (\Pi_1 - \Pi_0) \\
    &=\Pi(n) + \sigma \xi \Delta \Pi,
    \end{aligned}
\end{equation}
where $\Delta \Pi = \Pi_1 - \Pi_0$ denotes the payoff difference between the abundant ($\Pi_1$) and barren ($\Pi_0$) environmental states, and the noise is applied directly only to this term. Here, $R$ is the reward for mutual cooperation, $S$ is the sucker's payoff when a cooperator interacts with a defector, $T$ is the temptation to defect against a cooperator, and $P$ is the punishment for mutual defection. Following the theoretical approach developed in Ref.~\cite{wang2026Noiseinduced}, the temporal evolution of the cooperator fraction $x \in [0, 1]$ under replicator dynamics for games subject to noise is given by
\begin{equation}\label{eq:RDN}
    \dot{x} = s\,x(1-x)\Big[(\pi_C-\pi_D) + (0.5-x) s k^2(\sigma_C-\sigma_D)^2\Big],
\end{equation}
where $k=\sigma$, $s$ denotes selection intensity, and $\sigma_C$ ($\sigma_D$) denotes the payoff of a cooperator (defector) associated with the difference matrix $\Delta \Pi$ under the influence of noise. In Eq.~\eqref{eq:RDN}, random fluctuations are represented by deterministic terms in the thermodynamic limit (i.e., for an infinite population) \cite{wang2026Noiseinduced}, and their effects are the main focus of this work. Notably, this approach was derived under the weak-selection regime ($0 < s \ll 1$) \cite{wang2026Noiseinduced}; however, it remains qualitatively accurate in many cases of strong selection. 

\subsection{Individual-based model}

We perform Monte Carlo simulations of a birth-death process in a well-mixed population of $N$ individuals, denoted by $\mathcal{N} = \{1, 2, ..., N\}$. In each round, every individual interacts with each of the other $N-1$ individuals and receives a payoff
\begin{equation}
    \pi_i = \frac{1}{N-1}\Sigma_{j \in \mathcal{N} \setminus \{i\}} \pi(i, j),
\end{equation}
where $\pi(i, j)$ denotes the payoff received by individual $i$ when playing with individual $j$, as specified by Eq.~\eqref{eq:payoff_matrix_main}. The fitness of an individual is given by
\begin{equation}
    \eta_i = \exp(s \pi_i). 
\end{equation}
An individual is selected for reproduction, and its offspring replaces one of the other individuals.

The environmental state is updated according to
\begin{equation}\label{eq:mc_env_update}
    n_{\tau} = n_{\tau-1} + \varepsilon n_{\tau-1}(1-n_{\tau-1})[\theta x_{\tau-1} - (1-x_{\tau-1})]\Delta t, 
\end{equation}
where we set $\Delta t = 1/N$ between steps $\tau-1$ and $\tau$ to maintain a consistent timescale between strategy evolution and environmental evolution. After each update of the environmental state, the available resource is updated according to $\tilde{n}_{\tau} = n_{\tau} + \sigma \xi$. 

Importantly, to investigate nontrivial dynamics without the population becoming trapped in absorbing states due to stochastic updating, we introduce a single mutant whenever the entire population consists of individuals sharing the same strategy, either $C$ or $D$. We also restrict the environmental state to $n \in [\mu, 1-\mu]$, where $\mu = 10^{-7}$ is a small cutoff introduced to prevent the system from reaching the absorbing states at $n^*=0$ and $n^*=1$ under the discrete update rule in Eq.~\eqref{eq:mc_env_update}. These absorbing states are purely mathematical consequences of the discrete update rule and have no physical meaning.

\section{Results and discussion}

\begin{figure*}

    \includegraphics[width=\linewidth]{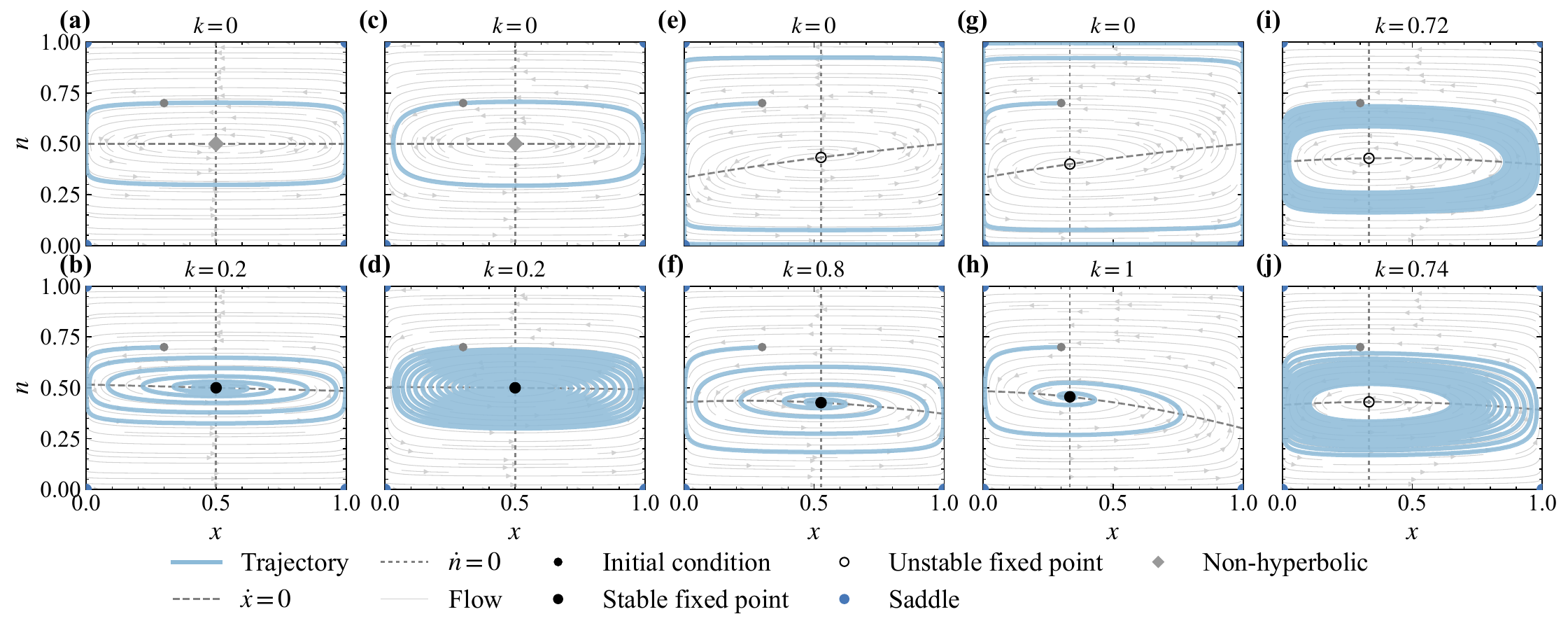}

  \caption{\label{figure_phase_diagram_ana} Analytical results and phase trajectories obtained by numerical integration. The phase diagrams of the coevolutionary dynamics of the cooperator fraction $x$ and environmental state $n$ are shown for the following cases: the donation game with $r=4$ for (a) $k=0$ and (b) $k=0.2$; the prisoner's dilemma with $R=3$, $S=0$, $T=5$ and $P=1$ for (c) $k=0$ and (d) $k=0.2$; the heteroclinic-cycle case with $\theta=0.9$ and $\Pi = \tilde{n}( (3, 0), (5, 2) )^\mathrm{T} + (1- \tilde{n}) ( (5, 1), (3, 0) )^\mathrm{T}$ for (e) $k=0$ and (f) $k=0.8$; and the heteroclinic-cycle case with $\theta=2$ for (g) $k=0$, (h) $k=1.0$, (i) $k=0.72$, and (j) $k=0.74$. Other parameters are $\varepsilon=0.01$, $\theta=1$ unless otherwise specified, and $s=0.1$. The initial condition $(0.3, 0.7)$ is chosen to facilitate visualization of the dynamical trajectory. } 
    
\end{figure*}

\begin{figure*}

    \includegraphics[width=\linewidth]{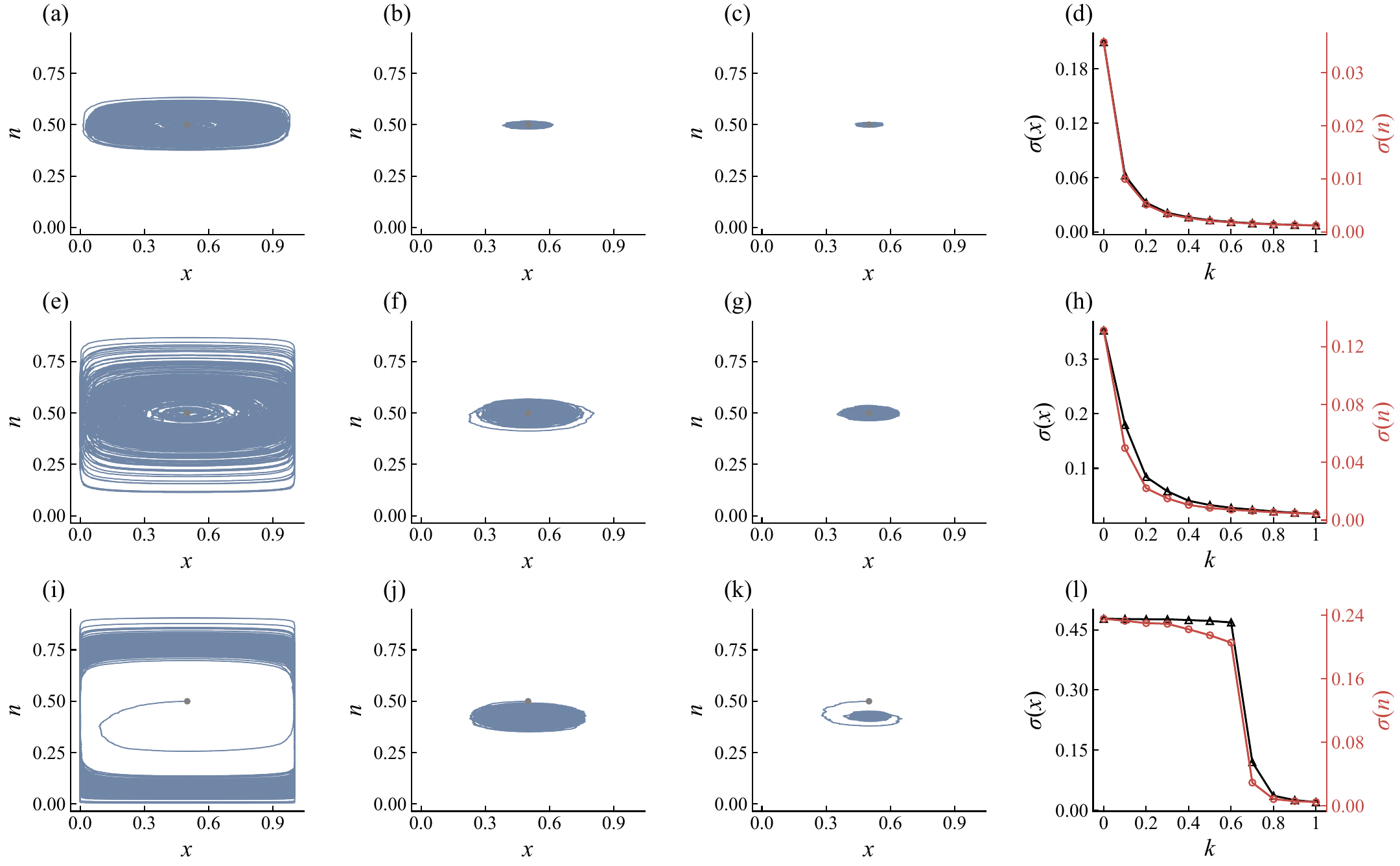}

  \caption{\label{figure_mc_phase_sigma} MC simulation results reveal that an increase in noise intensity $k$ generally leads to more stable system dynamics. The phase diagrams of the coevolutionary dynamics of the cooperator fraction $x$ and environmental state $n$ are shown for the following cases: the donation game with $r=4$ for (a) $k=0$, (b) $k=0.2$, and (c) $k=0.4$; the prisoner's dilemma with $R=3$, $S=0$, $T=5$, and $P=1$ for (e) $k=0$, (f) $k=0.2$, and (g) $k=0.4$; and the heteroclinic-cycle case with $\Pi = \tilde{n}( (3, 0), (5, 2) )^\mathrm{T} + (1- \tilde{n}) ( (5, 1), (3, 0) )^\mathrm{T}$ for (i) $k=0$, (j) $k=0.7$, and (k) $k=0.9$, respectively. Panels (d), (h) and (l) show the standard deviations $\sigma(x)$ and $\sigma(n)$ of the cooperator fraction $x$ and environmental state $n$, respectively, for the three cases. Other parameters are $\varepsilon=0.01$, $N=40000$, $\theta=1$, and $s=0.1$.  } 
    
\end{figure*}

\begin{figure*}
    \includegraphics[width=\textwidth]{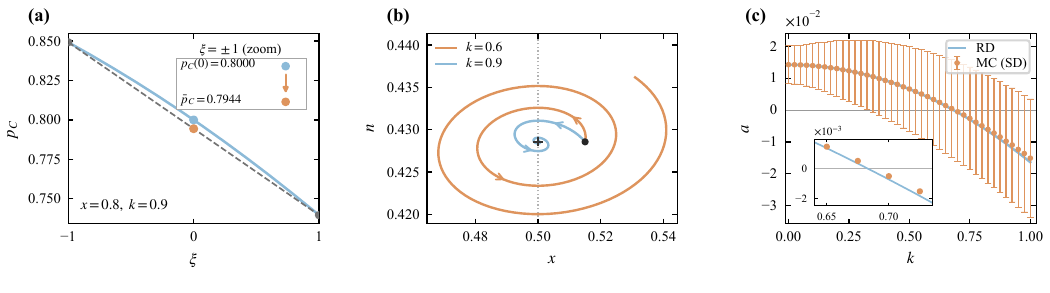}
    \caption{\label{figure_rectification_mechanism}
Rectification of unbiased payoff fluctuations leads to a stabilized interior fixed point for the payoff matrices in Eq.~\eqref{eq_pf_3052}. (a) Probability $p_C(\xi)$ of choosing a cooperator as the parent, as a function of the noise realization $\xi$ entering the payoffs through $n+k\xi$. Here $x=0.8$ and $k=0.9$, and $n$ is chosen so that the payoff difference between $C$ and $D$ vanishes at $\xi=0$. The points at $\xi=\pm1$ represent equal and opposite perturbations, and their equal-weight average $[p_C(-1)+p_C(1)]/2$ lies below the unperturbed probability $p_C(0)$, as shown in the inset. (b) Replicator dynamic trajectories for $k=0.6$ and $0.9$ starting from $(x_0,n_0)=(0.515,3/7)$. (c) Local Jacobian coefficient $a$ from replicator dynamics (RD, blue line) and Monte Carlo simulations (MC, orange points). Each MC point is obtained from the mean finite-difference response matrix over $2\times10^5$ independent experiments per $k$, with error bars showing one sample standard deviation of the individual estimates of $a$. The inset enlarges the region near $a=0$. Parameters are $N=40000$, $s=0.1$, $\varepsilon=0.01$, and $\theta=1$.}
\end{figure*}

\subsection{Fixed points\label{sec_rd_fp}}

The complete analysis is provided in the Appendices. Here, we present the essential steps needed to discuss the main results and conclusions.

For convenience, we define $\alpha_1 = R_1 - T_1$, $\alpha_0 = R_0 - T_0$, $\beta_1 = S_1 - P_1$ and $\beta_0 = S_0 - P_0$. We further define $\alpha(n) = \alpha_0 + n\alpha_\Delta$ and $\beta(n) = \beta_0 + n\beta_\Delta$, where $\alpha_\Delta = \alpha_1-\alpha_0$ and $\beta_\Delta = \beta_1-\beta_0$. We then rewrite Eqs.~\eqref{eq:te_env} and \eqref{eq:RDN} as
\begin{subequations}\label{eq:sim_rd_ed}
    \begin{align}
        \dot{x} &= s\,x(1-x)\,F(x,n),\\
        \dot{n} &= \varepsilon\,n(1-n)\,H(x),
    \end{align}
\end{subequations}
where $F(x,n) = \alpha(n)x+\beta(n)(1-x) + (0.5-x)s k^2\,D^2(x)$, $H(x) = \theta x-(1-x) = (1+\theta)x-1$ and $D(x) = \alpha_\Delta x + \beta_\Delta(1-x)$.

The four corner fixed points $(0,0),\;(1,0),\;(0,1),\;(1,1)$ exist for all parameter values. An interior fixed point $(x^*, n^*)$ exists when $H(x^*)=0$ and $F(x^*, n^*)=0$, with $x^*, n^* \in (0, 1)$. In addition, up to six non-corner boundary fixed points may exist, corresponding to solutions of $F(x_0^*, 0)=0$ and $F(x_1^*, 1)=0$, with $x_0^*, x_1^* \in (0, 1)$.

The corner fixed points $(1, 0)$ and $(0, 1)$ are unstable for all parameter values, as each has at least one positive eigenvalue of the Jacobian. The fixed point $(0, 0)$ is stable when $s\big(\beta_0+\frac{s k^2}{2}\beta_\Delta^2\big) < 0$, since its other eigenvalue, $-\varepsilon$, is always negative. Similarly, the fixed point $(1, 1)$ is stable when $s\big(\frac{s k^2}{2}\alpha_\Delta^2-\alpha_1\big) < 0$, since its other eigenvalue, $-\varepsilon \theta$, is always negative. The boundary fixed points $(x_0^*, 0)$ are stable when both $s x_0^*(1-x_0^*) F_{x}(x_0^*, 0) < 0$ and $\varepsilon H(x_0^*) < 0$ are satisfied. Likewise, the boundary fixed points $(x_1^*, 1)$ are stable when both $s x_1^*(1-x_1^*) F_{x}(x_1^*, 1) < 0$ and $-\varepsilon H(x_1^*) < 0$ are satisfied.

\paragraph*{Interior fixed point.}

In this paper, we focus on the effects of noise on o-TOC dynamics. Accordingly, our analysis focuses on the stability of the interior fixed point $(x^*, n^*)$. Solving the fixed-point conditions yields $x^* = \frac{1}{\theta+1}$ and $n^* = -\frac{\alpha_0+\theta \beta_0}{\alpha_\Delta+\theta \beta_\Delta} - sk^2\frac{(\theta-1)(\alpha_\Delta+\theta \beta_\Delta)}{2(\theta+1)^2}$. The condition $x^* = \frac{1}{\theta+1} \in (0, 1)$ is always satisfied for $\theta \in (0, +\infty)$. Thus, the existence of this fixed point requires $n^* \in (0, 1)$. Its stability is determined by the eigenvalues of the Jacobian evaluated at this fixed point
\begin{equation}\label{eq:J_xs_ns}
    J(x^*,n^*) = \begin{pmatrix}
    a & b\\[2pt]
    c & 0
    \end{pmatrix},
\end{equation}
where $a = s x^*(1-x^*) F_x^*$, $b = s x^*(1-x^*) D^*$, $c = \varepsilon n^*(1-n^*)(1+\theta)$ and $F_x^* \equiv F_x(x^*, n^*) \equiv \frac{\partial F}{\partial x}(x^*, n^*)$.

The eigenvalues satisfy $\lambda^2 - a\lambda - bc = 0$. Since $bc = \varepsilon s x^*(1-x^*) n^*(1-n^*)(1+\theta) D^*$, the sign of $bc$ is determined by the sign of $D^* = \frac{\alpha_\Delta + \theta \beta_\Delta}{\theta +1}$. The eigenvalues are given by
\begin{equation}\label{eq:jacobian_ifp_eigen}
    \lambda_\pm = \frac{1}{2}(a \pm \sqrt{a^2+4bc}). 
\end{equation}
Thus, complex eigenvalues occur when $a^2 + 4bc < 0$. For $D^* > 0$, the fixed point is necessarily unstable, since $a + \sqrt{a^2+4bc} > 0$ for any real value of $a$. For $D^* < 0$, the eigenvalues remain complex in a sufficiently small neighborhood of $a=0$, and the signs of their real parts are determined by the sign of $a$. Thus, the direction of the stability transition induced by varying $k$ depends on its effect on $a$. It is important to note that quantifying the strength of the progressive convergence requires a more detailed analysis. Here, we focus only on the direction of the noise-induced change in stability.

Recall that the term $s x^*(1-x^*)$ is always positive; therefore, the sign of $a$ is determined by the sign of $F_x^*$. We further derive that
\begin{equation}\label{eq:Fx_star_ana}
    \begin{aligned}
        F_x^* &= (\alpha_0 - \beta_0) - \frac{(\alpha_\Delta - \beta_\Delta)(\alpha_0 
        + \theta \beta_0)}{\alpha_\Delta + \theta \beta_\Delta} \\ 
        &\quad- \frac{1}{2}sk^2\frac{(\alpha_\Delta + \theta \beta_\Delta)[(3-\theta)\alpha_\Delta + (3\theta -1)\beta_\Delta]}{(\theta+1)^2},
    \end{aligned}
\end{equation}
and it is easy to see that $(\theta+1)^2 > 0$ and that $a$ decreases as $k$ increases for $k \geq 0$ when the sign of the term
\begin{equation}\label{eq:noise_led_stable}
    G_\mathrm{stable} = (\alpha_\Delta + \theta \beta_\Delta)[(3-\theta)\alpha_\Delta + (3\theta -1)\beta_\Delta]
\end{equation}
is positive. 
Thus, for $G_\mathrm{stable} > 0$, the stability transition at $a=0$ is always from unstable to stable as the noise intensity $k$ increases. This condition is always satisfied for $\theta=1$, whereas the case $\theta=-1$ is meaningless.

However, owing to the complexity of noise-induced replicator dynamics \cite{wang2026Noiseinduced}, a further systematic analysis of general cases with adjustable payoff elements would be difficult and is not necessary for the purposes of this work. Therefore, in the following sections, we present several informative and interesting cases and discuss the underlying mechanisms to support the central conclusion of this work. It is important to note that the Appendices provide a complete theoretical framework for analyzing arbitrary payoff structures whose elements are constant.

\subsection{Donation game\label{sec_rd_dg}}

We first consider a standard one-parameter donation game and its row-swapped form to illustrate the stability-enhancing effect of noise in this setting. The corresponding payoff matrices are defined as follows:
\begin{equation}
    \Pi^{\mathrm{DG}}_1 = \begin{bmatrix}
        5-r & -r \\
        5 & 0 \\    
    \end{bmatrix}, \quad
    \Pi^{\mathrm{DG}}_0 = \begin{bmatrix}
        5 & 0 \\ 
        5-r & -r \\   
    \end{bmatrix},
\end{equation}
where we choose $5$ (rather than the value $1$ commonly adopted in this literature) because a relatively large absolute value of $r$, compared with the other parameters, is beneficial for the subsequent analysis. Accordingly, $r \in (0, 5)$ quantifies the dilemma strength in the Prisoner's Dilemma game, and the dilemma becomes stronger as $r$ increases. Correspondingly, the payoff matrix for the barren environment, $\Pi^{\mathrm{DG}}_0$, is obtained by swapping the rows of $\Pi^{\mathrm{DG}}_1$, such that cooperators are favored under barren environmental conditions.

According to Eq.~\eqref{eq:jacobian_ifp_eigen}, the stability of the interior fixed point depends on $a$. We analytically derive that $a = s\frac{\theta}{(\theta+1)^2}(-4sk^2r^2) \leq 0$ for donation games, with equality only at $k=0$. Thus, in the absence of noise ($k=0$), the system exhibits a family of neutrally stable periodic trajectories. Upon introducing noise, the interior fixed point becomes stable. As shown in Figs.~\ref{figure_phase_diagram_ana}(a) and (b), the phase diagrams show periodic trajectories at $k=0$ and a stable interior fixed point at $k=0.2$, respectively. Due to the stochastic birth-death processes, the Monte Carlo simulations do not exhibit neutrally stable periodic trajectories (Fig.~\ref{figure_mc_phase_sigma}). However, as the noise level increases, fluctuations in the phase trajectories decrease, and the standard deviations of the statistics also decrease. Taken together, these results qualitatively confirm the conclusion that noise in resource availability promotes stability.

\subsection{Prisoner's dilemma\label{sec_rd_pd}}

Given the relationships among the entries of the payoff matrix in the Prisoner's Dilemma game, we further formulate a payoff structure involving this game and its row-swapped form. The payoff matrices are defined as follows:
\begin{equation}
    \Pi_0=\begin{bmatrix}T & P\\ R & S\end{bmatrix}, \quad
    \Pi_1=\begin{bmatrix}R & S\\ T & P\end{bmatrix},
\end{equation}
where $T>R>P>S$ and $2R>T+S$. According to Ref.~\cite{weitz2016oscillating} and the analysis in the Appendices, this class of games generally exhibits families of neutrally stable periodic trajectories in the absence of fluctuations in resource availability (i.e., $k=0$). Moreover, the non-corner boundary fixed points do not exist when $k=0$.

Interestingly, the result in Eq.~\eqref{eq:noise_led_stable} can be further simplified. For $\theta \in [\frac{1}{3}, 3]$, an increase in the noise intensity $k$ always leads to stable interior fixed points. For values of $\theta$ outside this interval, the condition $\frac{3\theta - 1}{\theta - 3} > \frac{R-T}{S-P}$ is required for $\theta >3$, whereas $\frac{3\theta - 1}{\theta - 3} < \frac{R-T}{S-P}$ is required for $\theta < \frac{1}{3}$.

To provide a complete analysis, we consider a specific case with the following payoff matrices:
\begin{equation}\label{eq_pf_3051}
    \Pi_0=\begin{bmatrix}5 & 1\\ 3 & 0\end{bmatrix}, \quad
    \Pi_1=\begin{bmatrix}3 & 0\\ 5 & 1\end{bmatrix}.
\end{equation}
As shown in Figs.~\ref{figure_phase_diagram_ana}(c) and (d), the phase diagrams show periodic trajectories at $k=0$ and a stable interior fixed point at $k=0.2$, respectively. The Monte Carlo simulations qualitatively reproduce these results (Fig.~\ref{figure_mc_phase_sigma}). As the noise intensity increases, the large-amplitude oscillatory trajectories give way to a state characterized by small-amplitude fluctuations arising from stochastic evolution.

\subsection{Heteroclinic cycle case\label{sec_rd_sc}}

In both preceding examples, noise damps oscillations that are neutrally stable in the absence of noise. We now investigate whether fluctuations can stabilize coexistence when the noiseless dynamics instead approach an attracting boundary heteroclinic cycle. Here, we analyze a simple modification of the case given by Eq.~\eqref{eq_pf_3051}. The payoff matrices are defined as follows:
\begin{equation}\label{eq_pf_3052}
    \Pi_0=\begin{bmatrix}5 & 1\\ 3 & 0\end{bmatrix}, \quad
    \Pi_1=\begin{bmatrix}3 & 0\\ 5 & 2\end{bmatrix},
\end{equation}
where we simply increase the element $P_1$. The $k=0$ case leads to a stable heteroclinic cycle. As the noise intensity $k$ associated with resource availability increases, the system undergoes a Hopf bifurcation, and the interior fixed point becomes stable upon reaching the transition point
\begin{equation}
k_C = \sqrt{\frac{4(\theta+1)^3}{s(3\theta+4)^2(5\theta+9)}}. 
\end{equation}
And the first Lyapunov coefficient 
\begin{equation}
        l_1 = \frac{s^2 k^2}{16} \frac{-66\theta^3-131\theta^2+47\theta+136}{(\theta+1)^2(3\theta+4)}
\end{equation}
determines whether the bifurcation is subcritical ($l_1 > 0$) or supercritical ($l_1 < 0$).

For $\theta = 0.9$, the Hopf bifurcation is subcritical, and increasing $k$ induces a dynamical transition from a heteroclinic cycle to a stable interior fixed point, as shown in Figs.~\ref{figure_phase_diagram_ana}(e) and (f). For $\theta = 2$, the Hopf bifurcation is supercritical, and stable limit cycles emerge for $k < k_C$. However, as $k$ decreases further below $k_C$, the limit-cycle orbit expands rapidly [Figs.~\ref{figure_phase_diagram_ana}(i) and (j)] and eventually merges with the boundary heteroclinic orbit [Fig.~\ref{figure_phase_diagram_ana}(g)]. For $k > k_C$, the system has a stable interior fixed point, as shown in Fig.~\ref{figure_phase_diagram_ana}(h).

The MC simulations for this case also qualitatively reproduce the conclusion that noise in resource availability promotes stable interior fixed points. However, in the Monte Carlo simulations, increasing the noise intensity always reduces the amplitude of the oscillatory trajectories, rendering heteroclinic-cycle and limit-cycle dynamics indistinguishable.

\subsection{Mechanism and discussion}

A microscopic analysis of the case in Eq.~\eqref{eq_pf_3052} illustrates how unbiased fluctuations in resource availability can induce a deterministic change in the dynamics, as shown in Fig.~\ref{figure_rectification_mechanism}. The nonlinear dependence of the cooperator-parent probability $p_C$ on the payoff difference rectifies unbiased payoff fluctuations into a shift in the average $p_C$ [Fig.~\ref{figure_rectification_mechanism}(a)]. In this example, the equal-weight average of $p_C$ at $\xi=\pm1$ (symmetrical noise events) is lower than its value in the absence of noise.

Thus, the nonlinear birth rule acts as a ratchet that transforms unbiased input into biased output. In the case considered here, increasing $k$ enhances the damping of oscillations and eventually stabilizes the interior fixed point through a Hopf bifurcation. Fig.~\ref{figure_rectification_mechanism}(b) illustrates the dynamics on either side of the transition. In particular, this mechanism effectively prevents the tragic o-TOC dynamics characterized by a heteroclinic cycle and maintains coexistence (as shown in Figs.~\ref{figure_phase_diagram_ana} and \ref{figure_mc_phase_sigma}). We also estimate the Jacobian coefficient $a$ at the interior fixed point from finite differences of the mean response measured in MC simulations. As shown in Fig.~\ref{figure_rectification_mechanism}(c), the MC simulation results estimates follow the trend predicted by replicator dynamics and capture the sign change of $a$ near the bifurcation. 

In this work, we have focused on specific cases because of the complexity of analyzing global stability for arbitrary payoff structures. Nevertheless, we have identified representative scenarios and elucidated the mechanisms by which fluctuations in resource availability can effectively alter o-TOC dynamics, particularly by promoting stable coexistence in the cases considered here. Furthermore, we provide a general theoretical framework in the appendices, which allows corresponding analyses under different scenarios by substituting specific parameter conditions. 

It is worth noting that the boundary heteroclinic cycle may coexist with other invariant structures in the interior of the phase space, including an interior fixed point or a limit cycle. In the parameter regimes considered here, however, the basin of attraction of the heteroclinic cycle is very small. Since such coexistence and the associated basin structure are beyond the scope of this work, we do not pursue their quantitative characterization.

We have employed Gaussian white noise in numerical simulations to examine the theoretical predictions. However, the original framework of noise-induced replicator dynamics is applicable to more complex scenarios \cite{wang2026Noiseinduced}. Future studies could apply the theoretical framework developed here to more complex and realistic ecological and social systems, as well as extend the present coevolutionary game dynamics to incorporate colored noise processes \cite{wang2026Noiseinduced,haunggi1994colored,jung1987dynamical,luczka2005nonmarkovian}.

\section{Conclusion}

In this work, we examine how fluctuations in resource availability alter evolutionary dynamics with environmental feedback. Motivated by the sensitivity of resource use to ambient conditions \cite{dahake2022Signallike,demanincor2026Climate,domeignoz-horta2020Microbial,marshall2021Wholeorganism,smith2021Systematic,zheng2019Growth,bacheler2020catchability}, we allow the effective availability of a shared resource to fluctuate at a given abundance, modeled as Gaussian noise. The nonlinear birth rule rectifies the resulting zero-mean payoff fluctuations, generating a deterministic correction to the replicator dynamics \cite{wang2026Noiseinduced,taylor1978evolutionarya}.

Our analysis shows how the payoff structure and feedback dynamics determine whether stronger fluctuations shift the interior equilibrium toward or away from local stability. In the examples studied, fluctuations damp neutrally stable oscillations and can stabilize coexistence even when the noiseless dynamics approach an attracting boundary cycle. Individual-based birth-death simulations corroborate the predicted reduction in the amplitudes of coupled strategy-resource oscillations. More broadly, these results reveal a mechanism by which fluctuations in resource availability can promote the coexistence of cooperation and defection and help sustain shared resources in systems prone to recurrent depletion, such as those exhibiting o-TOC dynamics.

\begin{acknowledgments}
This work was supported by the National Natural Science Foundation of China (Grants Nos.~12575039, 12375032 and 12247101), the Fundamental Research Funds for the Central Universities (Grant Nos.~lzujbky-2025-it50, lzujbky-2024-11, lzujbky-2024-jdzx06 and lzujbky-2023-ey02), the Natural Science Foundation of Gansu Province (No.~22JR5RA389), Gansu Provincial Department of Education: 2026 Graduate Innovation Star Project ``2026CXZX-153'', and the 111 Center under Grant No.~B20063. This work was partly supported by Longyuan-Youth-Talent Project of Gansu Province. 

We used AI tools (ChatGPT 5.6, 6.0, 6.1 and 6.2) to assist with code generation and review, language editing, and verification of mathematical derivations. We have reviewed all AI-generated content and take full responsibility for the content of this manuscript. 
\end{acknowledgments}

\section*{Data Availability Statement}
    All details required to reproduce the results of this paper are provided in the text. Further data are available upon reasonable request from authors.

\appendix

\begin{widetext}

\section{Fixed points and their existence}

The temporal evolution equations for $x$ and $n$ are given by \cite{wang2026Noiseinduced}:
\begin{equation}\label{eq:system}
\begin{aligned}
\dot{x} &= s\,x(1-x)\Big[(\pi_C-\pi_D) + (0.5-x)s k^2(\sigma_C-\sigma_D)^2\Big],\\
\dot{n} &= \varepsilon\,n(1-n)\big[\theta x-(1-x)\big],
\end{aligned}
\end{equation}
with the payoff matrices
\begin{equation}
    \Pi_0=\begin{bmatrix}R_0 & S_0\\ T_0 & P_0\end{bmatrix}, \quad
    \Pi_1=\begin{bmatrix}R_1 & S_1\\ T_1 & P_1\end{bmatrix},
\end{equation}
and the interpolated payoff matrix $\Pi$ and the corresponding matrix difference $\Delta \Pi$ are defined as $\Pi = (1-n)\Pi_0 + n \Pi_1$ and $\Delta \Pi = \Pi_1-\Pi_0$, respectively.

The payoffs of cooperators and defectors in the deterministic game, denoted by $\pi_C$ and $\pi_D$, respectively, are defined as follows:
\begin{equation}\label{eq:pi_payoff}
    \begin{aligned}
        \pi_C &= \Pi_{00}x+\Pi_{01}(1-x),\\ 
        \pi_D &= \Pi_{10}x+\Pi_{11}(1-x),
    \end{aligned}
\end{equation}
The payoffs of cooperators and defectors in games with fluctuations, denoted by $\sigma_C$ and $\sigma_D$, respectively, are defined as follows:
\begin{equation}\label{eq:sigma_payoff}
    \begin{aligned}
        \sigma_C = \Delta \Pi_{00}x+\Delta \Pi_{01}(1-x),\\
        \sigma_D = \Delta \Pi_{10}x+\Delta \Pi_{11}(1-x).
    \end{aligned}
\end{equation}
For convenience, we further define
\begin{equation}\label{eq:alpha_beta}
    \begin{aligned}
        \alpha_1 = R_1 - T_1,\quad \beta_1 = S_1 - P_1, \\
        \alpha_0 = R_0 - T_0,\quad \beta_0 = S_0 - P_0, \\
        \alpha_\Delta = \alpha_1-\alpha_0,\quad \beta_\Delta = \beta_1-\beta_0.
    \end{aligned}
\end{equation}
Combining Eqs.~\eqref{eq:pi_payoff} and \eqref{eq:sigma_payoff}, we obtain
\begin{equation}
    \begin{aligned}
        \pi_C-\pi_D &= \alpha(n)x + \beta(n)(1-x), \\
        \sigma_C-\sigma_D &= \alpha_\Delta x + \beta_\Delta(1-x) \equiv D(x),
    \end{aligned}
\end{equation}
where
\begin{equation}
    \alpha(n) = \alpha_0 + n\alpha_\Delta,\qquad \beta(n) = \beta_0 + n\beta_\Delta.
\end{equation}
Thus, the dynamical equations simplify to
\begin{subequations}\label{eq:compact}
    \begin{align}
        \dot{x} &= s\,x(1-x)\,F(x,n),\\
        \dot{n} &= \varepsilon\,n(1-n)\,H(x),
    \end{align}
\end{subequations}
with
\begin{subequations}\label{eq:F_H}
    \begin{align}
        F(x,n) &= \alpha(n)x+\beta(n)(1-x) + (0.5-x)s k^2\,D^2(x), \\
        H(x)   &= \theta x-(1-x) = (1+\theta)x-1, 
    \end{align}
\end{subequations}
where parameters \(s,k,\varepsilon,\theta\) are strictly positive.

Stationarity requires $\dot x=0$ and $\dot n=0$. From Eq.~\eqref{eq:compact}, $\dot{x} = 0$ implies $x\in\{0,1\}\;\text{or}\; F(x,n)=0$, while $\dot{n} = 0$ implies $n\in\{0,1\}\;\text{or}\; H(x)=0$. Moreover, $H(x)=0$ yields
\begin{equation}\label{eq:x_center}
    x^* = \frac{1}{1+\theta}. 
\end{equation}
Thus, the system has four corner fixed points, up to six non-corner boundary fixed points arising from the two cubic equations, and one interior fixed point. Their existence conditions are given as follows.

\subsection{Corner fixed points}
The four corner fixed points $(0,0),\;(1,0),\;(0,1),\;(1,1)$ exist for all parameter values.

\subsection{Non-corner boundary fixed points\label{sec_FP_BI}}
If $F(x,0)=0$ has roots $x_0^*\in(0,1)$, each root gives a boundary fixed point $(x_0^*,0)$. The corresponding equation is
\begin{equation}
    F(x_0^*, 0) = (\alpha_0 - \beta_0)x + \beta_0 + \frac{1}{2}s k^2 D^2(x) - s k^2 x D^2(x) = 0.
\end{equation}
Similarly, each root $x_1^*\in(0,1)$ of $F(x,1)=0$ gives a boundary fixed point $(x_1^*,1)$. The corresponding equation is
\begin{equation}
    F(x_1^*, 1) = (\alpha_1 - \beta_1)x + \beta_1 + \frac{1}{2}s k^2 D^2(x) - s k^2 x D^2(x) = 0.
\end{equation}
The existence of these fixed points depends on the parameter values and must therefore be checked case by case.

\subsection{Interior fixed point}
Using Eq.~\eqref{eq:x_center}, we set $x=x^*=\frac{1}{1+\theta}$. The fixed-point condition $F(x^*,n^*)=0$ then yields the following linear equation in $n^*$:
\begin{equation}
    \big[\alpha_0 x^*+\beta_0(1-x^*)\big] + n^*\big[\alpha_\Delta x^*+\beta_\Delta(1-x^*)\big] + (0.5-x^*)s k^2 [D(x^*)]^2 = 0.
\end{equation}
We further define
\begin{equation}
    D^* \equiv D(x^*) = \alpha_\Delta x^* + \beta_\Delta(1-x^*).
\end{equation}
For $D^*\neq 0$, we obtain
\begin{equation}\label{eq:nstar}
n^* = -\,\frac{\alpha_0 x^*+\beta_0(1-x^*) + (0.5-x^*)s k^2 (D^*)^2}{D^*} = -\frac{\alpha_0+\theta \beta_0}{\alpha_\Delta+\theta \beta_\Delta} - sk^2\frac{(\theta-1)(\alpha_\Delta+\theta \beta_\Delta)}{2(\theta+1)^2}.
\end{equation}
A feasible interior fixed point requires $n^*\in(0,1)$. For $D^*=0$, the fixed-point condition becomes degenerate, and this case is not considered further.

\section{Jacobian matrix of fixed points}
Let \(f(x,n)=s x(1-x)F(x,n)\) and \(g(x,n)=\varepsilon n(1-n)H(x)\). We adopt the notation \(\frac{\partial f}{\partial x}\equiv f_x\), with analogous definitions for $f_n, g_x, g_n, F_x, F_n$ and $H_x$. The partial derivatives are given by
\begin{subequations}\label{eq:f_g_xn}
    \begin{align}
        f_x &= s\big[(1-2x)F + x(1-x)F_x\big],\\
        f_n &= s x(1-x)F_n,\\
        g_x &= \varepsilon n(1-n)H_x(x) = \varepsilon n(1-n)(1+\theta),\\
        g_n &= \varepsilon\big[(1-2n)H(x)\big].
    \end{align}
\end{subequations}
Note that \(F_n(x,n) = D(x)\) and
\begin{equation}\label{eq:Fx}
F_x(x,n) = \alpha(n)-\beta(n) + s k^2\Big[-D(x)^2 + 2(0.5-x)D(x)(\alpha_\Delta-\beta_\Delta)\Big].
\end{equation}
All derivatives are evaluated at the fixed point under consideration.

\subsection{Jacobians of corner fixed points}

\begin{itemize}

\item \((0,0)\): \(F(0,0)=\beta_0+\frac12 s k^2\beta_\Delta^2\), \(H(0)=-1\).
\begin{equation}
    J(0,0)=\begin{pmatrix}
    s\big(\beta_0+\frac{s k^2}{2}\beta_\Delta^2\big) & 0\\[4pt]
    0 & -\varepsilon
    \end{pmatrix}.
\end{equation}
\item \((1,0)\): \(F(1,0)=\alpha_0-\frac12 s k^2\alpha_\Delta^2\), \(H(1)=\theta\).
\begin{equation}
    J(1,0)=\begin{pmatrix}
    s\big(\frac{s k^2}{2}\alpha_\Delta^2-\alpha_0\big) & 0\\[4pt]
    0 & \varepsilon\theta
    \end{pmatrix}.
\end{equation}
\item \((0,1)\): \(F(0,1)=\beta_1+\frac12 s k^2\beta_\Delta^2\), \(H(0)=-1\).
\begin{equation}
    J(0,1)=\begin{pmatrix}
    s\big(\beta_1+\frac{s k^2}{2}\beta_\Delta^2\big) & 0\\[4pt]
    0 & \varepsilon
    \end{pmatrix}.
\end{equation}
\item \((1,1)\): \(F(1,1)=\alpha_1-\frac12 s k^2\alpha_\Delta^2\), \(H(1)=\theta\).
\begin{equation}
    J(1,1)=\begin{pmatrix}
    s\big(\frac{s k^2}{2}\alpha_\Delta^2-\alpha_1\big) & 0\\[4pt]
    0 & -\varepsilon\theta
    \end{pmatrix}.
\end{equation}

\end{itemize}

\subsection{Jacobians of non-corner boundary fixed points}

\begin{itemize}

\item \((x_0^*,0)\): \(F(x_0^*,0)=0\), \(n=0\).  
\begin{equation}
    J(x_0^*,0)=\begin{pmatrix}
    s x_0^*(1-x_0^*) F_{x}(x_0^*, 0) & s x_0^*(1-x_0^*)D(x_0^*)\\[4pt]
    0 & \varepsilon H(x_0^*)
    \end{pmatrix}.
\end{equation}
where 
\begin{subequations}
    \begin{align}
        D(x_0^*) &= \alpha_\Delta x_0^* + \beta_\Delta (1- x_0^*), \\
        F_x(x_0^*, 0) &= \alpha_0 - \beta_0 + sk^2[-D^2(x_0^*)+(1-2x_0^*)D(x_0^*)(\alpha_\Delta-        \beta_\Delta)].
    \end{align}
\end{subequations}

\item \((x_1^*,1)\): \(F(x_1^*,1)=0\), \(n=1\).  
\begin{equation}
    J(x_1^*,1)=\begin{pmatrix}
    s x_1^*(1-x_1^*) F_{x}(x_1^*, 1) & s x_1^*(1-x_1^*)D(x_1^*)\\[4pt]
    0 & -\varepsilon H(x_1^*)
    \end{pmatrix}.
\end{equation}
where 
\begin{subequations}
    \begin{align}
        D(x_1^*) &= \alpha_\Delta x_1^* + \beta_\Delta (1- x_1^*), \\
        F_x(x_1^*, 1) &= \alpha_1 - \beta_1 + sk^2[-D^2(x_1^*)+(1-2x_1^*)D(x_1^*)(\alpha_\Delta-        \beta_\Delta)].
    \end{align}
\end{subequations}

\end{itemize}

Accordingly, the existence and stability of these fixed points must be examined under the corresponding parameter conditions.

\subsection{Jacobian of the interior fixed point}

At the interior fixed point $(x^*,n^*)$, we have $H(x^*)=0$ and $F(x^*,n^*)=0$. Consequently, $g_n=0$ and
\begin{equation}
    f_x = s x^*(1-x^*) F_x^*,\quad f_n = s x^*(1-x^*) D^*,\quad g_x = \varepsilon n^*(1-n^*)(1+\theta),
\end{equation}
where $F_x^* \equiv F_x(x^*, n^*)$ and $D^* \equiv D(x^*)$.
Thus,
\begin{equation}\label{eq:Jstar}
    J(x^*,n^*) = \begin{pmatrix}
    a & b\\[2pt]
    c & 0
    \end{pmatrix},
\end{equation}
where
\begin{subequations}
    \begin{align}
        a &= s x^*(1-x^*) F_x^*, \\
        b &= s x^*(1-x^*) D^*, \\
        c &= \varepsilon n^*(1-n^*)(1+\theta).
    \end{align}
\end{subequations}
The eigenvalues satisfy \(\lambda^2 - a\lambda - bc = 0\). Since \(bc = \varepsilon s x^*(1-x^*) n^*(1-n^*)(1+\theta) D^*\), the sign of \(bc\) equals the sign of \(D^*\). The eigenvalues are given by
\begin{equation}
    \lambda_\pm = \frac{1}{2}(a \pm \sqrt{a^2+4bc}). 
\end{equation}
Thus, complex eigenvalues arise when \(a^2 + 4bc < 0\).

\section{Boundary heteroclinic cycle}

In this section, we analyze the boundary heteroclinic cycle in a simple case in which no non-corner boundary fixed points, as discussed in Sec.~\ref{sec_FP_BI}, exist. The four edges of the unit square are invariant. The direction of flow along each edge is determined by the sign of the eigenvalue associated with the corresponding tangential direction at each corner fixed point. A simple heteroclinic cycle exists if all four corner fixed points are saddles, with stable and unstable directions alternating along the boundary, provided that no additional fixed points interrupt the connections along the boundary edges. The required signs are
\begin{equation}\label{eq:het_conditions}
\begin{aligned}
\beta_0 + \frac{s k^2}{2}\beta_\Delta^2 &> 0 \quad&\text{(unstable at }(0,0)\text{ along }n=0),\\[4pt]
\frac{s k^2}{2}\alpha_\Delta^2 - \alpha_0 &< 0 \quad&\text{(stable at }(1,0)\text{ along }n=0),\\[4pt]
\frac{s k^2}{2}\alpha_\Delta^2 - \alpha_1 &> 0 \quad&\text{(unstable at }(1,1)\text{ along }n=1),\\[4pt]
\beta_1 + \frac{s k^2}{2}\beta_\Delta^2 &< 0 \quad&\text{(stable at }(0,1)\text{ along }n=1).
\end{aligned}
\end{equation}
If, in addition, $F(x,0)\neq 0$ and $F(x,1)\neq 0$ for all $x\in(0,1)$, the boundary edges contain no non-corner fixed points, and the heteroclinic cycle is therefore unblocked. The orientation of the cycle is given by
\[
(0,0)\xrightarrow{\;n=0\;}(1,0)\xrightarrow{\;x=1\;}(1,1)\xrightarrow{\;n=1\;}(0,1)\xrightarrow{\;x=0\;}(0,0).
\]
It is important to note that the stability of the heteroclinic cycle requires further analysis under the specific conditions in which non-corner boundary fixed points exist.

\subsection{Stability of the heteroclinic cycle}
The stability of the cycle restricted to the invariant boundary is determined by the product of the ratios of the absolute values of the stable eigenvalues to those of the unstable eigenvalues at the four saddle points \cite{hofbauer1998evolutionarya}:
\begin{align}
\rho_{00} &= \frac{\varepsilon}{s(\beta_0+\frac12 s k^2\beta_\Delta^2)},\qquad
\rho_{10} = \frac{s(\alpha_0-\frac12 s k^2\alpha_\Delta^2)}{\varepsilon\theta},\nonumber\\
\rho_{11} &= \frac{\varepsilon\theta}{s(\frac12 s k^2\alpha_\Delta^2-\alpha_1)},\qquad
\rho_{01} = \frac{-s(\beta_1+\frac12 s k^2\beta_\Delta^2)}{\varepsilon}.
\end{align}
The overall stability index is given by
\begin{equation}\label{eq:Gamma}
        \Gamma = \rho_{00}\,\rho_{10}\,\rho_{11}\,\rho_{01}
        = \frac{(\alpha_0-\frac12 s k^2\alpha_\Delta^2)(\beta_1+\frac12 s k^2\beta_\Delta^2)} {(\beta_0+\frac12 s k^2\beta_\Delta^2)(\alpha_1-\frac12 s k^2\alpha_\Delta^2)} . 
\end{equation}

\begin{itemize}
\item \(\Gamma > 1\): the cycle is asymptotically stable on the boundary.
\item \(\Gamma = 1\): the cycle is at the critical threshold, and
    its stability cannot be determined from the eigenvalue
    ratios alone.
\item \(\Gamma < 1\): the cycle is unstable on the boundary.
\end{itemize}

It is important to note that $\Gamma>1$ guarantees only the
local asymptotic stability of the heteroclinic cycle within
the invariant boundary; it does not imply that the cycle
globally attracts interior initial conditions. This is because
interior fixed points may coexist with the cycle and alter
the global basin structure.

\section{Hopf bifurcation and stability of limit cycle}

\subsection{Critical condition}
A Hopf bifurcation at the interior fixed point \((x^*,n^*)\) requires a pair of purely imaginary eigenvalues. From Eq.~\eqref{eq:Jstar}, this occurs when
\begin{equation}\label{eq:hopf_cond}
a = s x^*(1-x^*) F_x^* = 0,\qquad\text{and}\qquad a^2+4bc < 0 \;\Longrightarrow\; bc < 0 \;\Longrightarrow\; D^* < 0.
\end{equation}
Expanding \(F_x^*\) yields the explicit condition for $ a = 0 $:
\begin{equation}\label{eq:hopf_cond_expand}
\alpha_0-\beta_0 + n^*(\alpha_\Delta-\beta_\Delta) + s k^2\Big[-(D^*)^2 + 2(0.5-x^*)D^*(\alpha_\Delta-\beta_\Delta)\Big] = 0.
\end{equation}

\subsection{First Lyapunov coefficient}

To determine the direction and stability of the bifurcating periodic orbit, we compute the first Lyapunov coefficient $l_1$. We shift the coordinate origin to the fixed point by defining $u=x-x^*$ and $v=n-n^*$. We then expand $(f,g)$ in a Taylor series about $(u,v)=(0,0)$. On the critical manifold, where $a=0$, the linear part is
$\begin{pmatrix}
0 & b\\
c & 0
\end{pmatrix}$.
We define $\omega = \sqrt{-bc}>0$. After a linear transformation,
\begin{equation}
    X = u,\qquad Y = \frac{b}{\omega}v,
\end{equation}
and after a linear transformation and rescaling time by $\omega$, the system becomes
\begin{equation}\label{eq:ode_XY}
    \dot X = \omega Y + \tilde P(X,Y),\qquad \dot Y = -\omega X + \tilde Q(X,Y).
\end{equation}
The standard formula for $l_1$ given in Ref.~\cite{guckenheimer1983nonlinear} (Section 3.4, Eq.~3.4.11) can be applied to the system described by Eq.~\eqref{eq:ode_XY}:
\begin{equation}\label{eq:GH}
\begin{aligned}
l_1 &= \frac{1}{16}\Big( \tilde P_{XXX} + \tilde P_{XYY} + \tilde Q_{XXY} + \tilde Q_{YYY} \Big)\\
&\quad + \frac{1}{16\omega}\Big[ - \tilde P_{XY}\big(\tilde P_{XX}+\tilde P_{YY}\big) + \tilde Q_{XY}\big(\tilde Q_{XX}+\tilde Q_{YY}\big) + \tilde P_{XX}\tilde Q_{XX} - \tilde P_{YY}\tilde Q_{YY} \Big],
\end{aligned}
\end{equation}
where the subscripts denote partial derivatives evaluated at $(0,0)$. Note that the minus sign appears in a different position from that in Ref.~\cite{guckenheimer1983nonlinear} because Eq.~\eqref{eq:ode_XY} has a different form.

In what follows, we apply this formula to the specific system considered in this work, exploiting its special structure to obtain a compact expression.

We expand $f$ and $g$ in Taylor series about $(0,0)$ up to third order. Using the facts that $F=0$, $H=0$, and $F_x=0$ at the fixed point, and noting that $H$ is linear in $x$, one can verify that many derivatives vanish. We denote the partial derivatives of $f$ and $g$ at the origin by
\[
f_{uu} = \frac{\partial^2 f}{\partial u^2},\quad
f_{uv} = \frac{\partial^2 f}{\partial u\partial v},\quad
f_{vv} = \frac{\partial^2 f}{\partial v^2},\quad
f_{uuu} = \frac{\partial^3 f}{\partial u^3},\quad
f_{uuv} = \frac{\partial^3 f}{\partial u^2\partial v},\quad
f_{uvv} = \frac{\partial^3 f}{\partial u\partial v^2},\quad
f_{vvv} = \frac{\partial^3 f}{\partial v^3},
\]
and analogously for $g$. Because of the coordinate transformation between $(x,n)$ and $(u,v)$, we have
$\frac{\partial f}{\partial u}\frac{\partial u}{\partial x} = \frac{\partial f}{\partial x}$,
and analogous relations hold for the corresponding derivatives of $f$ and $g$ with respect to $u$, $v$, $x$, and $n$. Thus, we further obtain:
\begin{equation}
    f_{vv}=0,\quad f_{vvv}=0,\quad f_{uvv}=0,\quad g_{uu}=0,\quad g_{vv}=0,\quad g_{uuu}=0,\quad g_{uuv}=0,\quad g_{vvv}=0.
\end{equation} 
Consequently, the Taylor expansions are given by
\begin{equation}\label{eq:taylor}
\begin{aligned}
\dot u &= b v + \frac12 f_{uu} u^2 + f_{uv} u v + \frac16 f_{uuu} u^3 + \frac12 f_{uuv} u^2 v + \frac12 f_{uvv} u v^2 + \cdots,\\
\dot v &= c u + g_{uv} u v + \frac12 g_{uvv} u v^2 + \cdots.
\end{aligned}
\end{equation}

Recall that $c = -\omega^2/b$.  Substituting into Eq.~\eqref{eq:taylor} gives
\begin{subequations}
    \begin{align}
        \nonumber \dot X &= b\left(\frac{\omega}{b} Y\right) + \frac12 f_{uu} X^2 + f_{uv} X\left(\frac{\omega}{b} Y\right) \\ 
        \nonumber &\qquad + \frac16 f_{uuu} X^3 + \frac12 f_{uuv} X^2\left(\frac{\omega}{b} Y\right) + \frac12 f_{uvv} X\left(\frac{\omega}{b} Y\right)^2 + \cdots\\
        &= \omega Y + \frac12 f_{uu} X^2 + \frac{\omega}{b} f_{uv} X Y + \frac16 f_{uuu} X^3 + \frac{\omega}{2b} f_{uuv} X^2 Y + \frac{\omega^2}{2b^2} f_{uvv} X Y^2 + \cdots, \\[4pt]
        \nonumber \dot Y &= \frac{b}{\omega}\,\dot v
        = \frac{b}{\omega}\left( c X + g_{uv} X\left(\frac{\omega}{b} Y\right) + \frac12 g_{uvv} X\left(\frac{\omega}{b} Y\right)^2 + \cdots\right)\\
        \nonumber &= \frac{b}{\omega}\left( -\frac{\omega^2}{b} X + \frac{\omega}{b} g_{uv} X Y + \frac{\omega^2}{2b^2} g_{uvv} X Y^2 + \cdots\right)\\
        &= -\omega X + g_{uv} X Y + \frac{\omega}{2b} g_{uvv} X Y^2 + \cdots.
    \end{align}
\end{subequations}

The system now takes the form of Eq.~\eqref{eq:ode_XY} with
\begin{equation}\label{eq:PQ}
    \begin{aligned}
        \tilde P(X,Y) &=
        \frac12 f_{uu} X^2 + \frac{\omega}{b} f_{uv} X Y + \frac16 f_{uuu} X^3 + \frac{\omega}{2b} f_{uuv} X^2 Y + \frac{\omega^2}{2b^2} f_{uvv} X Y^2, \\[4pt]
        \tilde Q(X,Y) &=
        g_{uv} X Y + \frac{\omega}{2b} g_{uvv} X Y^2, 
    \end{aligned}
\end{equation}
where higher-order terms are omitted.

From Eq.~\eqref{eq:PQ}, we compute all the required second- and third-order partial derivatives evaluated at $(X,Y)=(0,0)$.

\textbf{Second order:}
\[
\begin{aligned}
\tilde P_{XX} &= f_{uu}, &\quad \tilde P_{XY} &= \frac{\omega}{b} f_{uv}, &\quad \tilde P_{YY} &= 0,\\[4pt]
\tilde Q_{XX} &= 0, &\quad \tilde Q_{XY} &= g_{uv}, &\quad \tilde Q_{YY} &= 0.
\end{aligned}
\]

\textbf{Third order:}
\[
\begin{aligned}
\tilde P_{XXX} &= f_{uuu}, &\quad \tilde P_{XXY} &= \frac{\omega}{b} f_{uuv}, &\quad \tilde P_{XYY} &= 0, &\quad \tilde P_{YYY} &= 0,\\[4pt]
\tilde Q_{XXX} &= 0, &\quad \tilde Q_{XXY} &= 0, &\quad \tilde Q_{XYY} &= \frac{\omega}{b} g_{uvv}, &\quad \tilde Q_{YYY} &= 0.
\end{aligned}
\]

Thus, we have:
\begin{equation}\label{eq:l1_2o}
    \tilde P_{XXX} + \tilde P_{XYY} + \tilde Q_{XXY} + \tilde Q_{YYY}
    = f_{uuu},
\end{equation}
and:
\begin{equation}\label{eq:l1_3o}
    \begin{aligned}
        &-\tilde P_{XY}(\tilde P_{XX}+\tilde P_{YY}) + \tilde Q_{XY}(\tilde Q_{XX}+\tilde Q_{YY}) + \tilde P_{XX}\tilde Q_{XX} - \tilde P_{YY}\tilde Q_{YY}\\
        &\qquad = -\tilde P_{XY}\tilde P_{XX} = -\frac{\omega}{b} f_{uv} f_{uu} = -\frac{\omega}{b} f_{uv} f_{uu}.
    \end{aligned}
\end{equation}

According to Eqs.~\eqref{eq:GH}, \eqref{eq:l1_2o} and \eqref{eq:l1_3o}, we have:
\begin{equation}\label{eq:l1_final_0}
    \begin{aligned}
        l_1 &= \frac{1}{16}\left( f_{uuu} \right) + \frac{1}{16\omega}\left( -\frac{\omega}{b} f_{uv} f_{uu} \right) \\[4pt]
        & = \frac{1}{16}\left( f_{uuu} - \frac{1}{b} f_{uv} f_{uu} \right),
    \end{aligned}
\end{equation}

The remaining derivatives $f_{uu}, f_{uv}, f_{uuu}$ can be expressed in terms of $F$ and its partial derivatives evaluated at $(x^*,n^*)$. For brevity, we define \(Z = x^*(1-x^*)\) and \(Z' = 1-2x^*\), yielding:
\begin{align}
f_{xx} &= s Z F_{xx}^*,\\
f_{xn} &= s Z' D^* + s Z (\alpha_\Delta-\beta_\Delta),\\
f_{xxx} &= 3 s Z' F_{xx}^* + s Z F_{xxx}^*,
\end{align}
where
\begin{equation}\label{eq:fxx_fxxx}
    F_{xx}^* = s k^2\big[-4 D^* (\alpha_\Delta-\beta_\Delta) + 2(0.5-x^*)(\alpha_\Delta-\beta_\Delta)^2\big],\qquad
    F_{xxx}^* = -6 s k^2 (\alpha_\Delta-\beta_\Delta)^2.
\end{equation}
Recall that $D^*=\alpha_\Delta x^* + \beta_\Delta (1-x^*)$ and $b = s x^* (1-x^*) D^*$. Substituting these relations into Eq.~\eqref{eq:l1_final_0} and simplifying yields the following compact expression:
\begin{equation}\label{eq:l1_final_1}
    l_1 = \frac{s}{16}\left[\, 2Z'F_{xx}^*  - \frac{Z}{D^*}\phi F_{xx}^* + ZF_{xxx}^* \,\right],
\end{equation}
where $\phi = \alpha_\Delta-\beta_\Delta$.

Substituting the expressions given in Eq.~\eqref{eq:fxx_fxxx} into this expression, we obtain:
\begin{equation}\label{eq:l1_final}
        l_1 = \frac{s^2\,k^2}{16}\left[\, -\frac{Z\,Z'}{D^*}\phi^3 + 2(Z'^2-Z)\phi^2 - 8Z'D^*\phi \,\right].
\end{equation}
The sign of \(l_1\) determines the type of Hopf bifurcation:
\begin{itemize}
\item \(l_1 < 0\): supercritical Hopf bifurcation, in which a stable limit cycle emerges.
\item \(l_1 > 0\): subcritical Hopf bifurcation, in which an unstable limit cycle is created.
\end{itemize}

For the $k$-induced bifurcation considered in this work, the transversality condition, \(\partial a/\partial k \neq 0\), should be verified at the critical parameter value.


\end{widetext}

\end{document}